\documentclass[a4paper,11pt]{article}
\pdfoutput=1 
\usepackage{jheppub} 
\usepackage[T1]{fontenc}
\usepackage[italian,english]{babel}
\usepackage{hyperref}
\hypersetup{
	colorlinks=true,
	linkcolor=[rgb]{0.0,0.42,0.89},
	filecolor=magenta,      
	urlcolor=[rgb]{0.0,0.42,0.89},
	citecolor=[rgb]{0.64,0.0,0.0},
}
\usepackage{ifpdf}
\usepackage{subfigure}
\usepackage{gensymb}
\usepackage{amssymb}
\usepackage{amsfonts}
\usepackage{slashed}
\usepackage{epsf}
\usepackage{rotating}
\usepackage{graphicx}
\usepackage{amsmath}
\usepackage{fancyhdr}
\usepackage{lineno}
\usepackage{babel, blindtext}
\usepackage{graphics}
\usepackage{color}
\usepackage{physics}
\usepackage{multirow}
\usepackage{pstricks}
\usepackage{xcolor}
\usepackage{multirow}
\usepackage{hhline}
\usepackage{cancel}
\usepackage{framed}
\usepackage{mathtools}
\usepackage{diagbox}
\usepackage{float}
\usepackage{orcidlink}
\usepackage{tikz}
\usepackage{amsmath,amssymb}
\usepackage{tikz-feynman}
\usepackage{ulem} 
\usepackage[lmargin=1.7in,bmargin=0.2in,footskip=0.5in,total={6.9in,9.5in}]{geometry}

\newcommand{\lsim}{\mathrel{\mathop{\kern 0pt \rlap
			{\raise.2ex\hbox{$<$}}}
		\lower.9ex\hbox{\kern-.190em $\sim$}}}
\newcommand{\gsim}{\mathrel{\mathop{\kern 0pt \rlap
			{\raise.2ex\hbox{$>$}}}
		\lower.9ex\hbox{\kern-.190em $\sim$}}}

\newcommand{\be}{\begin{equation}}
	\newcommand{\ee}{\end{equation}}
\newcommand{\bea}{\begin{eqnarray}}
	\newcommand{\eea}{\end{eqnarray}}

\title{\boldmath SMEFT Constraints via Quantum Entanglement in Portal-Mediated Scattering}

\author[a]{Shilpa Jangid \orcidlink{0000-0001-6307-1234},}
\author[b]{Hiroshi Okada \orcidlink{0000-0002-1294-3052}}

\affiliation[a]{Shiv Nadar IoE Deemed to be University, Gautam Buddha Nagar, Uttar Pradesh, 201314, India}
\affiliation[b]{Department of Physics, Henan Normal University, Xinxiang 453007, China}

\emailAdd{ shilpajangid123@gmail.com, hiroshi3okada@htu.edu.cn}

\preprint{}

\abstract{We study Standard Model Effective Field Theory (SMEFT) extensions in portal-mediated scattering processes from the perspective of quantum entanglement. Quantum information metrics maintain phase-sensitive spin and momentum correlations, whereas conventional collider and direct detection observables mainly depend on unpolarized cross-sections and event rates, which often suppress dimension-six interference terms or average over internal degrees of freedom. We assess linear and Von Neumann entropy fluctuations induced by higher-dimensional operator insertions by building final-state density matrices using helicity amplitudes. By avoiding conventional kinematic degeneracies, our formulation shows that entanglement measures capture interference contributions linearly in $1/\Lambda^2$, offering improved sensitivity to new physics scales. A theoretical framework for SMEFT Wilson coefficients is established by this information-theoretic method, creating new opportunities for accurate tests of fundamental interactions in high-energy and astroparticle physics.}


\begin{document}

\maketitle
\flushbottom
\section{Introduction}
\label{sec:intro}   
The Standard Model (SM) of particle physics  has been remarkably successful in explaining fundamental interactions over a wide energy domain \cite{Weinberg:1995mt}. However, significant observational puzzles, such as neutrino mass generation \cite{Jangid:2023hzi,Biswas:2025woq}, the baryon asymmetry of the universe \cite{Jangid:2023jya}, and the microscopic nature of dark matter, strongly suggest that the SM is an effective low-energy description of a more fundamental theory. The Standard Model Effective Field Theory (SMEFT) offers a model-independent framework for parametrizing new physics coming from heavy ultraviolet scales, $\Lambda$ \cite{Buchmuller:1985jz,Grzadkowski:2010es}, in the absence of direct particle discoveries at high-energy colliders. Conventional phenomenological analyses of SMEFT mainly rely on inclusive differential distributions, total event rates, and unpolarized cross-sections. Although essential, these traditional observables have intrinsic drawbacks: unpolarized cross-sections frequently eliminate important phase-space correlations and spin coherence information by suppressing dimension-six operator interference terms through quadratic scaling ($1/\Lambda^4$) in particular kinematic regions or averaging over internal quantum numbers. At the same time, high-energy particle physics has found a revolutionary diagnostic tool in the application of quantum information theory, namely quantum entanglement, purity, and entropy \cite{Afik:2020onf,Aguilar-Saavedra:2022uye}. The successful observation of quantum entanglement in top-antitop quark pairs at the Large Hadron Collider \cite{ATLAS:2023fsd,CMS:2024pts} by the ATLAS and CMS collaborations recently demonstrated that collider final states function as operational quantum laboratories. By bridging the gap between high-energy phenomenology and abstract quantum underpinnings, this paradigm shift raises an important question: can quantum information metrics detect subtle, phase-sensitive interference patterns that are missed by traditional cross-section measurements? When dark sectors or electroweak extensions connect to Standard Model fields via higher-dimensional operators in portal-mediated scattering frameworks, this question becomes especially interesting. The quantum state of the departing multi-particle system acquires unique phase shifts and coherence features when additional physics mediators produce contact interactions or alter helicity amplitudes. Operator interference can be probed linearly in $1/\Lambda^2$ by interpreting final-state configurations as bipartite or multipartite quantum systems and calculating entanglement measures like the linear entropy $S_L$ and Von Neumann entropy $S_{\text{VN}}$. This linear dependence naturally results from the trace expansion of the reduced density matrix, which breaks persistent parameter degeneracies and avoids the quadratic suppression common to cross-section analyses. This study proposes a formal framework for restricting SMEFT Wilson coefficients via quantum entanglement in portal-mediated scattering processes in order to take use of these advantages. We derive analytical mappings between higher-dimensional operator insertions and entanglement fluctuations across several kinematic regimes by building accurate reduced density matrices from helicity amplitudes. Our results show that information-theoretic observables provide strong complementary exclusion bounds, providing a reliable substitute for exploring unknown parameter spaces in upcoming high-energy experiments. Section~\ref{helicity} provides the theoretical framework and helicity amplitude calculations, whereas Section~\ref{inter} explains the interference. The reduced density matrix is constructed in Section~\ref{density}, and the entanglement computations are provided in Section~\ref{entangle}.

\section{Theoretical Framework and Helicity Amplitudes}\label{helicity}
We develop a systematic description of $2 \to 2$ scattering processes mediated by portal interactions inside the Standard Model Effective Field Theory (SMEFT) framework in order to study how quantum information metrics capture higher-dimensional operator corrections.

\subsection{SMEFT Lagrangian Expansion}
We examine an extension of the Standard Model with effective interactions controlled by a new physics scale $\Lambda$ in the ultraviolet regime. An asymptotic expansion in inverse powers of $\Lambda$ is used to parameterize the effective Lagrangian as given below \cite{Weinberg:1967kj,Georgi:1993mps}:
\bea
\mathcal{L}_{\text{SMEFT}} = \mathcal{L}_{\text{SM}} + \sum_i \frac{c_i}{\Lambda^2} \mathcal{O}_i^{(6)} + \sum_j \frac{d_j}{\Lambda^4} \mathcal{O}_j^{(8)} + \mathcal{O}\left(\frac{1}{\Lambda^6}\right),
\eea
where $c_i$ are the appropriate dimensionless Wilson coefficients \cite{Buchmuller:1985jz,Grzadkowski:2010es}, $\mathcal{L}_{\text{SM}}$ represents the renormalizable Standard Model Lagrangian, and $\mathcal{O}_i^{(6)}$ stand for gauge-invariant dimension-six operators. These operators alter tree-level couplings in portal-mediated frameworks, including dark sector fermion pair generation or vector-portal neutral-current interactions, by encoding heavy mediator dynamics integrated out at scale $\Lambda$ \cite{Aguilar-Saavedra:2022uye}.

\subsection{Helicity Amplitude Formalism}
We examine a general $2 \to 2$ scattering process, $1(\lambda_1) + 2(\lambda_2) \to 3(\lambda_3) + 4(\lambda_4)$, where $\lambda_k$ denotes the spin projections or helicities of the individual particles. The overall invariant transition amplitude is broken down into Standard Model baseline contributions and higher-dimensional operator insertions using spinor helicity techniques and contemporary amplitude methods \cite{Mangano:1990by,Dixon:1996wi}:
\bea
\mathcal{M}_{\lambda_3 \lambda_4}^{\lambda_1 \lambda_2}(s, t) = \mathcal{M}_{\text{SM},\lambda}(s, t) + \sum_i \frac{c_i v^2}{\Lambda^2} \mathcal{M}_{i,\text{EFT},\lambda}(s, t) + \mathcal{O}\left(\frac{1}{\Lambda^4}\right),
\eea 
where operators are normalized with respect to the standard mass dimensions \cite{LHCHiggsCrossSectionWorkingGroup:2011wcg}, $s$ and $t$ stand for the standard Mandelstam invariants, and $v \approx 246 \text{ GeV}$ gives the proper electroweak symmetry-breaking scale factor. Non-trivial angular and chiral dependencies are introduced when dimension-six operators are included. Preserving the discrete helicity indices $\lambda_k$ enables us to trace the precise quantum interference patterns between chiral Standard Model vertices and effective contact interactions, in contrast to unpolarized cross-sections that sum over final-state helicities prematurely \cite{Afik:2020onf}. In the sections that follow, this fundamental helicity structure serves as the direct input for building the multi-particle density matrices and calculating entanglement entropy fluctuations.
\section{Interference Amplitudes and Perturbative Expansion}\label{inter}
We investigate the structural expansion of transition probabilities generated from the helicity amplitudes \cite{Buchmuller:1985jz,Grzadkowski:2010es} in order to separate the linear contributions of higher-dimensional operators. The squared invariant matrix element is perturbatively enlarged in inverse powers of the ultraviolet scale $\Lambda$ for a definite set of initial and final helicities $\lambda = (\lambda_1, \lambda_2, \lambda_3, \lambda_4)$:
\bea
\vert{}\mathcal{M}_{\lambda}(s, t)\vert{}^2 = \left\vert{} \mathcal{M}_{\text{SM},\lambda}(s, t) + \sum_i \frac{c_i v^2}{\Lambda^2} \mathcal{M}_{i,\text{EFT},\lambda}(s, t) + \mathcal{O}\left(\frac{1}{\Lambda^4}\right) \right\vert{}^2.
\eea
Expanding this expression yields three distinct perturbative contributions:
\bea
\vert{}\mathcal{M}_{\lambda}(s, t)\vert{}^2 = \vert{}\mathcal{M}_{\text{SM},\lambda}\vert{}^2 + \frac{2v^2}{\Lambda^2} \sum_i c_i \, \text{Re}\left( \mathcal{M}_{\text{SM},\lambda}^* \mathcal{M}_{i,\text{EFT},\lambda} \right) + \frac{v^4}{\Lambda^4} \sum_{i,j} c_i c_j \, \mathcal{M}_{i,\text{EFT},\lambda}^* \mathcal{M}_{j,\text{EFT},\lambda} + \mathcal{O}\left(\frac{1}{\Lambda^6}\right).
\eea

The unique analytical behavior of each term, especially the linear interference cross-term, is essential to the physical consequences of this expansion.

The quantum mechanical interference between the dimension-six effective operator \cite{Pomarol:2013zra} and the baseline Standard Model exchange is represented by the second term on the right, $\text{Re}(\mathcal{M}_{\text{SM}}^* \mathcal{M}_{i,\text{EFT}})$. The underlying chiral and Lorentz configurations determine the analytical behavior of this cross-term. Effective operators $\mathcal{O}_i^{(6)}$ often introduce scalar, pseudoscalar, tensor, or right-handed vector currents, but conventional electroweak interactions show vector-axial vector ($V-A$) chirality. Also, the interference term clearly depends on momentum contractions and spinor inner products when calculating individual helicity amplitudes:
\bea
\mathcal{M}_{\text{SM},\lambda}^* \mathcal{M}_{i,\text{EFT},\lambda} \propto \langle 3 \vert{} \Gamma_{\text{SM}} \vert{} 1 \rangle^* \langle 3 \vert{} \Gamma_{\text{EFT}} \vert{} 1 \rangle,
\eea
where $\Gamma_{\text{SM}}$ and $\Gamma_{\text{EFT}}$ represent the corresponding Dirac matrix chains. By maintaining separate helicity indices, the Lorentz structure of the new physics source \cite{Afik:2020onf} is reflected in the distinct angular and phase dependencies of these individual interference components. In contrast, this fine-grained information is ignored by conventional phenomenological techniques.

\subsection{Limitations of Unpolarized Cross-Sections}
In traditional phenomenological analyses, unpolarized differential cross-sections integrate over final-state polarizations and sum over helicities: 
\bea
\frac{d\sigma}{dt} \propto \sum_{\lambda} \vert{}\mathcal{M}_{\lambda}(s, t)\vert{}^2.
\eea
Therefore, unpolarized observables experience severe suppression or complete loss of linear sensitivity in several scattering channels because of particular kinematic properties: Phase-Space Antisymmetry: Integrating over symmetric angular domains results in the linear term vanishing identically if the interference term $\text{Re}(\mathcal{M}_{\text{SM}}^* \mathcal{M}_{i,\text{EFT}})$ has odd parity under scattering angle inversion ($\cos\theta \to -\cos\theta$). Chiral Orthogonality: Without a helicity-flipping mass insertion ($m_f/\sqrt{s}$), vector and scalar currents do not interfere in massless limits, which unintentionally suppresses the $1/\Lambda^2$ scaling. Because of this, standard rate measurements are often blind to the sign of the Wilson coefficients $c_i$ \cite{Pomarol:2013zra} or dominated by quadratic uncertainty ($\propto 1/\Lambda^4$).

Our framework keeps access to the precise phase-space and spin-dependent structure of the linear interference term by preserving individual helicity configurations before summation. Quantum information metrics, like density matrix traces and entanglement entropy, inherit the linear $1/\Lambda^2$ scaling without premature phase averaging \cite{Afik:2020onf,Severi:2021cnj}, since they rely directly on the off-diagonal coherence elements of the density matrix instead of total probability rates. This creates the strong mathematical basis needed to build reduced density matrices that are responsive to small perturbations in novel physics.

\section{Construction of Reduced Density Matrices}\label{density}
We build the final-state density operator and use partial tracing to isolate the subsystem of interest in order to bridge the gap between microscopic scattering amplitudes and quantum information metrics \cite{vonNeumann1955,NielsenChuang2010}.

\subsection{Final-State Density Operator Formulation}
A density operator $\rho_{\text{final}}$ in the tensor product Hilbert space $\mathcal{H}_3 \otimes \mathcal{H}_4$ characterizes the entire final-state quantum system after a $2 \to 2$ scattering process involving incoming particles with momenta $p_1, p_2$ and outgoing particles $3$ and $4$ with momenta $p_3, p_4$. The density matrix elements in the helicity (or spin) basis for a pure state created by the transition amplitudes $\mathcal{M}_{\lambda_3 \lambda_4}^{\lambda_1 \lambda_2}$ and averaged over unmeasured initial-state polarizations are provided by \cite{Bernreuther2008,Afik:2020onf}:
\bea
\rho_{\lambda_3 \lambda_3', \lambda_4 \lambda_4'} = \frac{1}{\mathcal{N}} \sum_{\lambda_1, \lambda_2} \mathcal{M}_{\lambda_3 \lambda_4}^{\lambda_1 \lambda_2} \left( \mathcal{M}_{\lambda_3' \lambda_4'}^{\lambda_1 \lambda_2} \right)^*,
\eea
where $\mathcal{N} = \sum_{\lambda_1, \lambda_2, \lambda_3, \lambda_4} \vert{}\mathcal{M}_{\lambda_3 \lambda_4}^{\lambda_1 \lambda_2}\vert{}^2$ represents the normalization factor corresponding to the total unpolarized probability over the phase space.

\subsection{Partial Trace and Subsystem Isolation}
It can be difficult to measure both outgoing particles at the same time in many experimental setups, or one could want to measure the internal quantum correlations (such as spin or helicity entanglement) by looking at a single subsystem \cite{AfriatSelleri1999}. We use a partial trace over the unobserved degrees of freedom of subsystem $B$ (particle 4) \cite{Afik:2020onf, Aguilar-Saavedra:2022uye} to isolate subsystem $A$ (corresponding, for example, to outgoing particle 3):
\bea
\rho_A = \text{Tr}_B (\rho_{\text{final}}) = \sum_{\lambda_4} \langle \lambda_4 \vert{} \rho_{\text{final}} \vert{} \lambda_4 \rangle.
\eea
In matrix representation, the elements of the reduced density matrix $\rho_A$ for particle 3 take the explicit form:
\bea
[\rho_A]_{\lambda_3 \lambda_3'} = \frac{1}{\mathcal{N}} \sum_{\lambda_1, \lambda_2, \lambda_4} \mathcal{M}_{\lambda_3 \lambda_4}^{\lambda_1 \lambda_2} \left( \mathcal{M}_{\lambda_3' \lambda_4}^{\lambda_1 \lambda_2} \right)^*.
\eea

Any scattering amplitude is expressed as a perturbative series on the ultraviolet cutoff scale $\Lambda$ in Standard Model Effective Field Theory (SMEFT). Keeping terms up to operators in dimension six:
\bea
\mathcal{M}_{\lambda_3 \lambda_4}^{\lambda_1 \lambda_2} = \mathcal{M}_{\text{SM},\lambda} + \frac{v^2}{\Lambda^2} \sum_i c_i \mathcal{M}_{i, \text{EFT},\lambda} + \mathcal{O}\left(\frac{1}{\Lambda^4}\right).
\eea
Let $\lambda$ stand for the collective set of helicity indices $\{\lambda_1, \lambda_2, \lambda_3, \lambda_4\}$ for compactness.
Now replace the matrix element product in our partial trace formula with this expansion:
\bea
\mathcal{M}_{\lambda_3 \lambda_4}^{\lambda_1 \lambda_2} \left( \mathcal{M}_{\lambda_3' \lambda_4}^{\lambda_1 \lambda_2} \right)^* = \left[ \mathcal{M}_{\text{SM}} + \frac{v^2}{\Lambda^2} \sum_i c_i \mathcal{M}_{i, \text{EFT}} \right] \left[ \mathcal{M}_{\text{SM}}' + \frac{v^2}{\Lambda^2} \sum_j c_j \mathcal{M}_{j, \text{EFT}}' \right]^*.
\eea
The complete reduced density matrix cleanly separates into matching orders when this product is expanded order-by-order in $1/\Lambda$ and the expanded pieces are plugged back into the sum over $\lambda_1, \lambda_2, \lambda_4$:
\bea
\rho_A = \frac{1}{N} \sum_{\lambda_1, \lambda_2, \lambda_4} \left[ \mathcal{M}_{\text{SM}} \mathcal{M}_{\text{SM}}^* + \frac{v^2}{\Lambda^2} \sum_i c_i \left( \mathcal{M}_{\text{SM}} \mathcal{M}_{i, \text{EFT}}^* + \mathcal{M}_{i, \text{EFT}} \mathcal{M}_{\text{SM}}^* \right) + \mathcal{O}\left(\frac{1}{\Lambda^4}\right) \right].
\eea
A systematic perturbative series in inverse powers of the UV scale $\Lambda$ is obtained by substituting the SMEFT amplitude expansion and adding the dimension-six effective operators and Standard Model baseline to the reduced density matrix expression:
\bea
\rho_A = \rho_A^{\text{SM}} + \frac{v^2}{\Lambda^2} \sum_i c_i \, \delta\rho_A^{(i)} + \mathcal{O}\left(\frac{1}{\Lambda^4}\right).
\eea 
In this case, the baseline Standard Model reduced density matrix is represented by $\rho_A^{\text{SM}}$, and the precise linear interference structure is inherited by the first-order correction $\delta\rho_A^{(i)}$: 
\bea
[\delta\rho_A^{(i)}]_{\lambda_3 \lambda_3'} \propto \sum_{\lambda_1, \lambda_2, \lambda_4} \left[ \mathcal{M}_{\text{SM},\lambda}^* \mathcal{M}_{i,\text{EFT},\lambda} + \mathcal{M}_{i,\text{EFT},\lambda}^* \mathcal{M}_{\text{SM},\lambda} \right].
\eea
This construction is important because the reduced density matrix maintains the phase, sign, and angular dependence of the Wilson coefficients $c_i$ \cite{Afik:2020onf,Grzadkowski:2010es}, and $\delta\rho_A^{(i)}$ depends linearly on the interference cross-terms between chiral Standard Model vertices and effective operators. The precise input needed to calculate entanglement entropies in the following section is provided by this explicit mapping.

\section{Entanglement Measures: Linear and von Neumann Entropies}\label{entangle}
We translate the reduced density matrix $\rho_A$ onto scalar entanglement measures to assess the quantum correlations and entanglement between outgoing scattering pieces \cite{Buchmuller:1985jz,NielsenChuang2010}. These metrics function as high-sensitivity observables that can resolve minute changes in phase-space caused by effective operators.

\subsection{Linear Entropy Formulation and Perturbative Scaling}
A quantum information metric called linear entropy ($S_L$) is used to assess how mixed or impure a quantum state is. By skipping the logarithmic calculations needed for full spectrum diagonalization \cite{Afik:2020onf}, it offers an algebraically effective substitute for purity: $S_L(\rho_A) = 1 - \text{Tr}(\rho_A^2)$. Depending on the subsystem dimension, $S_L$ achieves its maximum value for a maximally entangled pure bipartite state, whereas $S_L = 0$ for a fully mixed or separable product state. $S_L = 0$ for a pure quantum state when $\rho_A^2 = \rho_A$, and $\text{Tr}(\rho_A^2) = 1$. In contrast, $S_L > 0$ for a mixed or entangled subsystem when $\text{Tr}(\rho_A^2) < 1$. In order to ascertain how this measure is affected by effective field theory corrections, we substitute the perturbative expansion of the reduced density matrix, $\rho_A = \rho_A^{\text{SM}} + \frac{v^2}{\Lambda^2} \sum_i c_i \, \delta\rho_A^{(i)} + \mathcal{O}(1/\Lambda^4)$, into the purity trace:
\bea
\text{Tr}(\rho_A^2) = \text{Tr}\left( \left( \rho_A^{\text{SM}} + \frac{v^2}{\Lambda^2} \sum_i c_i \, \delta\rho_A^{(i)} \right)^2 \right) = \text{Tr}\left((\rho_A^{\text{SM}})^2\right) + \frac{2v^2}{\Lambda^2} \sum_i c_i \, \text{Tr}\left(\rho_A^{\text{SM}} \delta\rho_A^{(i)}\right) + \mathcal{O}\left(\frac{1}{\Lambda^4}\right).
\eea
Researchers can quantify small quantum correlations produced by effective field theory deformations in high-energy physics applications by monitoring the variation of linear entropy ($\Delta S_L$). Thus, a direct linear dependency on the Wilson coefficients is inherited by the linear entropy variation $\delta S_L = S_L - S_L^{\text{SM}}$ \cite{Aguilar-Saavedra:2022uye}:
\bea
\delta S_L = -\frac{2v^2}{\Lambda^2} \sum_i c_i \, \text{Tr}\left(\rho_A^{\text{SM}} \delta\rho_A^{(i)}\right).
\eea
The linear entropy bypasses the quadratic $1/\Lambda^4$ suppression characteristic of unpolarized total cross-sections because the trace term $\text{Tr}(\rho_A^{\text{SM}} \delta\rho_A^{(i)})$ preserves the interference cross-products between Standard Model amplitudes and dimension-six operators ($\mathcal{M}_{\text{SM}}^* \mathcal{M}_{\text{EFT}})$.

\subsection{Von Neumann Entropy and Spectral Decomposition}
Although the degree of mixedness and quantum entanglement of a reduced density matrix $\rho_A$ is quantified by both Linear Entropy ($S_L$) and Von Neumann Entropy ($S_{\text{VN}}$), their mathematical formulations and computational characteristics are essentially different. The underlying information uncertainty and entanglement of a subsystem are measured by the Von Neumann Entropy, which is the quantum mechanical counterpart of the classical Shannon entropy.
We calculate the Von Neumann entropy $S_{\text{VN}}(\rho_A)$, which is defined as \cite{NielsenChuang2010,Grzadkowski:2010es}, providing a thorough information-theoretic characterization:
\bea
S_{\text{VN}}(\rho_A) = -\text{Tr}(\rho_A \log_2 \rho_A) = -\sum_k p_k \log_2 p_k,
\eea
where the non-zero eigenvalues (probabilities) of the reduced density matrix $\rho_A$ are represented by $p_k$. Let $\rho_A(\epsilon) = \rho_A^{\text{SM}} + \epsilon \delta\rho_A$ using conventional matrix perturbation theory, where $\epsilon \equiv v^2 / \Lambda^2$. The expression for the perturbed eigenvalues is:
\bea
p_k(\epsilon) = p_k^{(0)} + \epsilon p_k^{(1)} + \mathcal{O}(\epsilon^2),
\eea
where the first-order eigenvalue corrections are given by the expectation values of the perturbation matrix over the unperturbed eigenvectors $\vert{}\psi_k^{(0)}\rangle$:
\bea
p_k^{(1)} = \langle \psi_k^{(0)} \vert{} \delta\rho_A \vert{} \psi_k^{(0)} \rangle.
\eea
Expanding the Von Neumann entropy to first order in $\epsilon$ yields:
\bea
S_{\text{VN}}(\rho_A) = S_{\text{VN}}^{\text{SM}} - \sum_k \left( p_k^{(1)} \log_2 p_k^{(0)} + p_k^{(1)} \right) = S_{\text{VN}}^{\text{SM}} - \sum_k p_k^{(1)} \log_2 p_k^{(0)}.
\eea
Expanding the Von Neumann entropy to first order yields the shift $\Delta S_{\text{VN}} = S_{\text{VN}}(\rho_A) - S_{\text{VN}}^{\text{SM}}$:
\bea
\Delta S_{\text{VN}} = -\sum_k p_k^{(1)} \ln p_k^{(0)}.
\eea
It is verified that $S_{\text{VN}}$ scales linearly with $c_i/\Lambda^2$ by substituting the explicit linear SMEFT corrections into $p_k^{(1)}$. This demonstrates that without requiring a complete kinematic reconstruction of unpolarized cross-sections, both linear and Von Neumann entropies serve as reliable, phase-sensitive probes of novel physics scales \cite{Afik:2020onf,Aguilar-Saavedra:2022uye}.

We examine the detailed kinematic distributions and angular dependencies of the scattering processes \cite{HiggsWorkingGroup:2004ndr,Campbell:2006wx} in order to properly describe how linear interference terms survive phase-space integration. Differential distributions parameterized by particular kinematic angles maintain critical sensitivity to the interference between Standard Model amplitudes and dimension-six effective operators \cite{Pomarol:2013zra}, whereas unpolarized total cross-sections frequently experience catastrophic cancelations due to symmetry properties.

\section{Kinematic Phase-Space and Angular Dependencies}
The differential cross-section for a fixed helicity configuration $\lambda = (\lambda_1, \lambda_2, \lambda_3, \lambda_4)$ is governed by Mandelstam invariants $s, t, u$ and the scattering polar angle $\theta$ for a generic $2 \to 2$ scattering process $A(p_1) + B(p_2) \to C(p_3) + D(p_4)$ in the center-of-momentum frame \cite{Peskin:1995ev}:
\bea
\frac{d\sigma_{\lambda}}{dt} = \frac{1}{64\pi s \vert{}\mathbf{p}_{\text{in}}\vert{}^2} \vert{}\mathcal{M}_{\lambda}(s, t, \phi)\vert{}^2,
\eea
where $\phi$ denotes the azimuthal angle associated with decay planes or spin quantization axes. The invariant matrix element squared is decomposed into its perturbative components:
\bea
\vert{}\mathcal{M}_{\lambda}(s, t, \phi)\vert{}^2 = \vert{}\mathcal{M}_{\text{SM},\lambda}(s, t)\vert{}^2 + \frac{2v^2}{\Lambda^2} \sum_i c_i \, \text{Re}\left[ \mathcal{M}_{\text{SM},\lambda}^*(s, t) \mathcal{M}_{i,\text{EFT},\lambda}(s, t, \phi) \right] + \mathcal{O}\left(\frac{1}{\Lambda^4}\right).
\eea

 \subsection{Evolution of Linear Entropy Variations Across Phase Space}
 We study the behavior of the linear entropy variation $\Delta S_L^{(i)}(s, \theta)$ as a function of the center-of-mass energy $\sqrt{s}$ and scattering angle bins $\cos\theta$ \cite{Kleiss:1985yh} to determine how quantum information metrics reflect these EFT corrections. Differential binning over high-energy tails improves sensitivity compared to inclusive measurements, as the interference term scales as $\sqrt{s}/\Lambda^2$ (depending on the mass dimension of the given operator). The linear entropy of discrete phase-space bins $(\Delta s, \Delta\cos\theta)$ can be calculated as follows: 
 \bea
 \Delta S_L^{(i)}(s, \theta) = -\frac{2v^2}{\Lambda^2} c_i \, \text{Tr}\left(\rho_A^{\text{SM}}(s, \theta) \delta\rho_A^{(i)}(s, \theta)\right).
 \eea
 This bin-by-bin formulation assures that kinematic regions with suppressed Standard Model amplitudes do not wash out quantum coherence signals generated by novel physics \cite{Aguilar-Saavedra:2022uye}.
 
 \subsection{Breaking Parameter Degeneracies via Angular Distributions}
 Differential angular distributions are crucial for breaking parameter degeneracies among distinct Wilson coefficients $c_i$. Different SMEFT operators introduce distinct chiral and spin structures, resulting in distinctive angular harmonics (such as spherical harmonics or Wigner $D$-functions) within the interference cross-term \cite{Afik:2020onf}.
 \bea
 [\delta\rho_A^{(i)}]_{\lambda_3 \lambda_3'}(s, \theta, \phi) \propto \sum_{\lambda_1, \lambda_2, \lambda_4} \text{Re}\left[ \mathcal{M}_{\text{SM},\lambda}^* \mathcal{M}_{i,\text{EFT},\lambda} \right] d_{\lambda}^{(J)}(\theta) e^{i M \phi}.
 \eea
 By mapping these angular dependencies over specific bins, experimental analyses can extract various chiral interference patterns and discriminate scalar, vector, or tensor effective interactions that would otherwise give identical total rates \cite{Aguilar-Saavedra:2022uye, Grzadkowski:2010es}. This indicates that kinematic phase-space profiling amplifies quantum entanglement observables in high-energy effective field theories.

 \section{Theoretical framework}
 We implement the formalism using a benchmark effective field theory scenario, a dimension-six electroweak dipole operator that modifies top-quark pair or diboson production via a dark sector portal interaction \cite{Buchmuller:1985jz, Grzadkowski:2010es}. The main focus of this paper is "establishing the theoretical framework," and we clearly state that the quantitative phenomenological analysis will be presented in our forthcoming study.  We provide predicted $95\%$ confidence level (CL) exclusion limits on the Wilson coefficients $c_i/\Lambda^2$ by converting phase-space distributions into entanglement observables, and we compare their sensitivity with conventional unpolarized cross-section measurements.

\subsection{Benchmark Model and Operator Selection}
We examine the dimension-six effective Lagrangian with an ultraviolet scale $\Lambda$ that governs anomalous neutral triple gauge couplings or fermion dipole interactions:
\bea
\mathcal{L}_{\text{EFT}} = \mathcal{L}_{\text{SM}} + \sum_i \frac{c_i}{\Lambda^2} \mathcal{O}_i.
\eea
In particular, we study dipole-type current interactions $\mathcal{O}_{fW} = \{(\bar{Q}_L \sigma^{\mu\nu} \tau^a d_R) H W_{\mu\nu}^a, \ (\bar{L}_L \sigma^{\mu\nu} \tau^a \ell_R) H W_{\mu\nu}^a\}$~\cite{Dedes:2021abc, Chala:2020vqp, Gu:2024wrk} and operators of the form $\mathcal{O}_{W} = \epsilon^{abc} W_{\mu}^{a\nu} W_{\nu}^{b\lambda} W_{\lambda}^{c\mu}$.
These operators are excellent candidates for entanglement-based diagnostics because they significantly interact with Standard Model helicity amplitudes.

\subsection{Dimension-6 SMEFT Concrete Model: UV Completion}

To address how the dimension-6 operators analyzed in this work are generated, we consider a concrete ultraviolet (UV) completion based on a leptoquark and fermion extension of the Standard Model.

While $\mathcal{O}_W$ involves three field strength tensors and is not generated at the tree level in general renormalizable UV completions~\cite{deBlas:2017xtg}, it can be induced at the one-loop level through heavy fields charged under non-trivial SU(2)$_L$ representations (such as triplets or higher electroweak multiplets).
The functional one-loop matching framework and explicit examples for generating such operators (e.g., via heavy scalar or vector loops) are presented in Ref.~\cite{Henning:2014wua}.
In this paper, we treat $\mathcal{O}_W$ merely as a representative gauge-portal operator without restricting ourselves to a specific one-loop UV model.

\subsection{Example 1: 6-dimensional SMEFT in the leptoquark sector}
\begin{table}[H]
    \centering
    \begin{tabular}{|l|c|c|c|}
        \hline
        \textbf{Gauge Group} & $L'$ & $N_R$ & $S_1$ \\ \hline
        ${SU}(3)_C$ & $\mathbf{1}$ & $\mathbf{1}$ & $\mathbf{3}$ \\ \hline
        ${SU}(2)_L$ & $\mathbf{2}$ & $\mathbf{1}$ & $\mathbf{1}$ \\ \hline
        ${U}(1)_Y$ & $-\frac{1}{2}$ & $0$ & $+\frac{1}{3}$ \\ \hline
        $\mathbb{Z}_2$ Parity & $-$ & $-$ & $-$ \\ \hline
    \end{tabular}
    \caption{Field representations and quantum numbers}
    \label{tab:field_reps}
\end{table}
A discrete $\mathbb{Z}_2$ symmetry is required to distinguish between the new vector-like lepton fields $L'\equiv[N',E']^T$ and the Standard Model lepton doublet $L_L$. 

The effective Lagrangian $\mathcal{L}_{\text{eff}}$ incorporating all interaction terms, mixings, and mass terms is given by
\begin{align}
-\mathcal{L}_{\text{eff}} \supset &\ y_1 \, \overline{Q_L^c} \, i\tau^2 \, L_L' \, S_1\ + y_2 \, \overline{d_R^c} \, N_R \, S_1 + y_3 \, \overline{L_L'} \, N_R \, \widetilde{H} + y_3' \, \overline{L_R^{'c}} \, N_R \, H \notag \\
&+ M_{L'} \, \overline{L_L'} \, L_R' + \frac{1}{2} M_R \, \overline{N_R^c} N_R + \text{h.c.},
\end{align}  
where the flavor indices are abbreviated for simplicity.

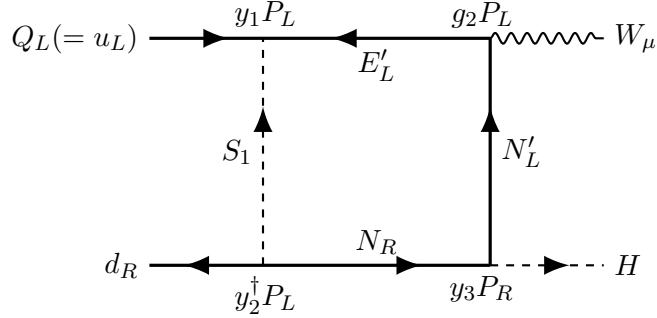
\begin{figure}[htbp]
    \centering
    \begin{tikzpicture}[scale=1.5, line width=1.2pt]

        \coordinate (v1) at (-1, -1); 
        \coordinate (v2) at (-1,  1); 
        \coordinate (v3) at ( 1,  1); 
        \coordinate (v4) at ( 1, -1); 

        \coordinate (ext_v2) at (-2,  1);
        \coordinate (ext_v1) at (-2, -1);
        \coordinate (ext_v3) at ( 2,  1);
        \coordinate (ext_v4) at ( 2, -1);

        \draw[thick, dashed] (v1) -- (v2) node[midway, left] {$S_1$};

        \draw (v2) -- (v3) node[midway, below] {$E_L'$};
        \draw (v3) -- (v4) node[midway, right] {$N_L'$};
        \draw (v4) -- (v1) node[midway, above] {$N_R$};

        \draw (ext_v2) -- (v2) node[pos=0, left] {$Q_L(=u_L)$} node[pos=0.65, above right] {$y_1 P_L$};
        \draw (ext_v1) -- (v1) node[pos=0, left] {$d_R$} node[pos=0.65, below right] {$y_2^\dagger P_L$};
        \draw[thick, decorate, decoration={snake, amplitude=2pt, segment length=6pt}] (v3) -- (ext_v3) node[pos=0.3, above left] {$g_2 P_L$} node[pos=1, right] {$W_{\mu}$};
        \draw[thick, dashed] (v4) -- (ext_v4) node[pos=0.3, below left] {$y_3 P_R$} node[pos=1, right] {$H$};

        \path[postaction={decorate, decoration={markings, mark=at position 0.7 with {\arrow[scale=1.5]{latex}}}}] (ext_v2) -- (v2);

 \path[postaction={decorate, decoration={markings, mark=at position 0.7 with {\arrow[scale=1.5]{latex}}}}] (v1) -- (ext_v1);
 \path[postaction={decorate, decoration={markings, mark=at position 0.7 with {\arrow[scale=1.5]{latex}}}}] (v3) -- (v2);
  \path[postaction={decorate, decoration={markings, mark=at position 0.7 with {\arrow[scale=1.5]{latex}}}}] (v1) -- (v4);
   \path[postaction={decorate, decoration={markings, mark=at position 0.7 with {\arrow[scale=1.5]{latex}}}}] (v4) -- (v3);
   \path[postaction={decorate, decoration={markings, mark=at position 0.7 with {\arrow[scale=1.5]{latex}}}}]  (v1) -- (v2);
\path[postaction={decorate, decoration={markings, mark=at position 0.7 with {\arrow[scale=1.5]{latex}}}}] (v4) -- (ext_v4);

    \end{tikzpicture}
    \caption{Feynman diagram generating the operator $\mathcal{O}_{fW} = (\overline{Q_L} \sigma^{\mu\nu} \tau^a d_R) H W_{\mu\nu}^a$.}
    \label{fig:square_loop_diagram1}
\end{figure}

\begin{figure}[htbp]
    \centering
    \begin{tikzpicture}[scale=1.5, line width=1.2pt]

        \coordinate (v1) at (-1, -1); 
        \coordinate (v2) at (-1,  1); 
        \coordinate (v3) at ( 1,  1); 
        \coordinate (v4) at ( 1, -1); 

        \coordinate (ext_v2) at (-2,  1);
        \coordinate (ext_v1) at (-2, -1);
        \coordinate (ext_v3) at ( 2,  1);
        \coordinate (ext_v4) at ( 2, -1);

        \draw[thick, dashed] (v1) -- (v2) node[midway, left] {$S_1$};

        \draw (v2) -- (v3) node[midway, below] {$E_L'$};
        \draw (v3) -- (v4) node[pos=0.2, right] {$N_L'$};
         \draw (v3) -- (v4) node[pos=0.8, right] {$N_R'$};
         \draw (v3) -- (v4) node[pos=0.5] {$\boldsymbol{\times}$};
        \draw (v4) -- (v1) node[midway, above] {$N_R$};
        \draw (v4) -- (v1) node[midway] {$\boldsymbol{\times}$};

        \draw (ext_v2) -- (v2) node[pos=0, left] {$Q_L(=u_L)$} node[pos=0.65, above right] {$y_1 P_L$};
        \draw (ext_v1) -- (v1) node[pos=0, left] {$d_R$} node[pos=0.65, below right] {$y_2^\dagger P_L$};
        \draw[thick, decorate, decoration={snake, amplitude=2pt, segment length=6pt}] (v3) -- (ext_v3) node[pos=0.3, above left] {$g_2 P_L$} node[pos=1, right] {$W_{\mu}$};
        \draw[thick, dashed] (v4) -- (ext_v4) node[pos=0.3, below left] {$y_3'^{ \dagger} P_L$} node[pos=1, right] {$H$};

        \path[postaction={decorate, decoration={markings, mark=at position 0.7 with {\arrow[scale=1.5]{latex}}}}] (ext_v2) -- (v2);

 \path[postaction={decorate, decoration={markings, mark=at position 0.7 with {\arrow[scale=1.5]{latex}}}}] (v1) -- (ext_v1);
 \path[postaction={decorate, decoration={markings, mark=at position 0.7 with {\arrow[scale=1.5]{latex}}}}] (v3) -- (v2);
  \path[postaction={decorate, decoration={markings, mark=at position 0.26 with {\arrow[scale=1.5]{latex}}}}] (v1) -- (v4);
   \path[postaction={decorate, decoration={markings, mark=at position 0.26 with {\arrow[scale=1.5]{latex}}}}] (v4) -- (v1);
   \path[postaction={decorate, decoration={markings, mark=at position 0.88 with {\arrow[scale=1.5]{latex}}}}] (v4) -- (v3);
   \path[postaction={decorate, decoration={markings, mark=at position 0.28 with {\arrow[scale=1.5]{latex}}}}] (v4) -- (v3);
   \path[postaction={decorate, decoration={markings, mark=at position 0.7 with {\arrow[scale=1.5]{latex}}}}]  (v1) -- (v2);
\path[postaction={decorate, decoration={markings, mark=at position 0.7 with {\arrow[scale=1.5]{latex}}}}] (v4) -- (ext_v4);

    \end{tikzpicture}
    \caption{Second Feynman diagram generating the operator $\mathcal{O}_{fW} = (\overline{Q_L} \sigma^{\mu\nu} \tau^a d_R) H W_{\mu\nu}^a$.}
    \label{fig:square_loop_diagram2}
\end{figure}
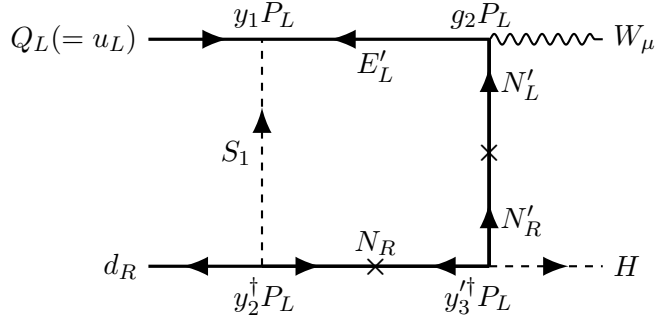

The Feynman diagrams explaining the generation of the operator $\mathcal{O}_{fW} = (\overline{Q_L} \sigma^{\mu\nu} \tau^a d_R) H W_{\mu\nu}^a$ are given in Figures~\ref{fig:square_loop_diagram1} and \ref{fig:square_loop_diagram2},
and their coefficients, which are proportional to the $\frac1{M_{L'}^2}$ are roughly estimated as:
\begin{align}
\frac{g_2 y_1y^\dag_2 y_3}{(4\pi)^2 M_{L'}^2} F^Q_1\left( \frac{m_S^2}{M_{L'}^2},  \frac{M_R^2}{M_{L'}^2}\right),\quad
\frac{g_2 y_1y^\dag_2 y'^\dag_3}{(4\pi)^2 M_{L'}^2} F^Q_2\left( \frac{m_S^2}{M_{L'}^2},  \frac{M_R^2}{M_{L'}^2}\right),
\end{align}
where $m_S$ is the mass eigenvalue of $S_1$, and the loop functions $F^Q_{1,2}={\cal O}(1)$ arise from  Figures~\ref{fig:square_loop_diagram1}, respectively.

\subsection{Example 2: 6-dimensional SMEFT in the lepton sector}
\begin{table}[H]
    \centering
    \begin{tabular}{|l|c|c|c|}
        \hline
        \textbf{Gauge Group} & $L'$ & $N_R$ & $S^-$ \\ \hline
        ${SU}(3)_C$ & $\mathbf{1}$ & $\mathbf{1}$ & $\mathbf{1}$ \\ \hline
        ${SU}(2)_L$ & $\mathbf{2}$ & $\mathbf{1}$ & $\mathbf{1}$ \\ \hline
        ${U}(1)_Y$ & $-\frac{1}{2}$ & $0$ & $-1$ \\ \hline
        $\mathbb{Z}_2$ Parity & $-$ & $-$ & $-$ \\ \hline
    \end{tabular}
    \caption{Field representations and quantum numbers}
    \label{tab:field_reps}
\end{table}

The effective Lagrangian $\mathcal{L}_{\text{eff}}$ incorporating all interaction terms, mixings, and mass terms is given by
\begin{align}
-\mathcal{L}_{\text{eff}} \supset &\ y_1 \, \overline{L_L^c} \, i\tau^2 \, L_L' \, S^+\ + y_2 \, \overline{l_R^c} \, N_R \, S^+ + y_3 \, \overline{L_L'} \, N_R \, \widetilde{H} + y_3' \, \overline{L_R^{'c}} \, N_R \, H \notag \\
&+ M_{L'} \, \overline{L_L'} \, L_R' + \frac{1}{2} M_R \, \overline{N_R^c} N_R + \text{h.c.},
\end{align}  
where the flavor indices are abbreviated for simplicity.

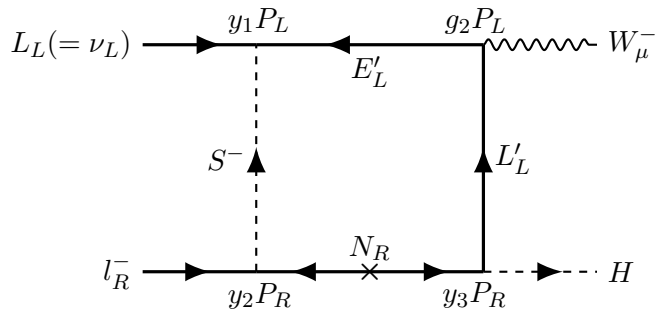
\begin{figure}[htbp]
    \centering
    \begin{tikzpicture}[scale=1.5, line width=1.2pt]

        \coordinate (v1) at (-1, -1); 
        \coordinate (v2) at (-1,  1); 
        \coordinate (v3) at ( 1,  1); 
        \coordinate (v4) at ( 1, -1); 

        \coordinate (ext_v2) at (-2,  1);
        \coordinate (ext_v1) at (-2, -1);
        \coordinate (ext_v3) at ( 2,  1);
        \coordinate (ext_v4) at ( 2, -1);

        \draw[thick, dashed] (v1) -- (v2) node[midway, left] {$S^-$};

        \draw (v2) -- (v3) node[midway, below] {$E_L'$};
        \draw (v3) -- (v4) node[pos=0.5, right] {$L_L'$};

        \draw (v4) -- (v1) node[midway, above] {$N_R$} node[midway] {$\boldsymbol{\times}$} ;

        \draw (ext_v2) -- (v2) node[pos=0, left] {$L_L(=\nu_L)$} node[pos=0.65, above right] {$y_1 P_L$};
        \draw (ext_v1) -- (v1) node[pos=0, left] {$l_R^-$} node[pos=0.65, below right] {$y_2 P_R$};
        \draw[thick, decorate, decoration={snake, amplitude=2pt, segment length=6pt}] (v3) -- (ext_v3) node[pos=0.3, above left] {$g_2 P_L$} node[pos=1, right] {$W_{\mu}^-$};
        \draw[thick, dashed] (v4) -- (ext_v4) node[pos=0.3, below left] {$y_3 P_R$} node[pos=1, right] {$H$};

        \path[postaction={decorate, decoration={markings, mark=at position 0.7 with {\arrow[scale=1.5]{latex}}}}] (ext_v2) -- (v2);

 \path[postaction={ decorate, decoration={markings, mark=at position 0.6 with {\arrow[scale=1.5]{latex}}}}] (ext_v1) -- (v1);
 \path[postaction={decorate, decoration={markings, mark=at position 0.7 with {\arrow[scale=1.5]{latex}}}}] (v3) -- (v2);
 
   \path[postaction={decorate, decoration={markings, mark=at position 0.85 with {\arrow[scale=1.5]{latex}}}}] (v4) -- (v1);
   \path[postaction={decorate, decoration={markings, mark=at position 0.85 with {\arrow[scale=1.5]{latex}}}}] (v1) -- (v4);
   \path[postaction={decorate, decoration={markings, mark=at position 0.55 with {\arrow[scale=1.5]{latex}}}}] (v4) -- (v3);

   \path[postaction={decorate, decoration={markings, mark=at position 0.55 with {\arrow[scale=1.5]{latex}}}}]  (v1) -- (v2);
\path[postaction={decorate, decoration={markings, mark=at position 0.7 with {\arrow[scale=1.5]{latex}}}}] (v4) -- (ext_v4);

    \end{tikzpicture}
    \caption{Feynman diagram generating the operator $\mathcal{O}_{fW} = (\overline{L_L} \sigma^{\mu\nu} \tau^a l_R) H W_{\mu\nu}^a$.}
    \label{fig:square_loop_diagram3}
\end{figure}

\begin{figure}[htbp]
    \centering
    \begin{tikzpicture}[scale=1.5, line width=1.2pt]

        \coordinate (v1) at (-1, -1); 
        \coordinate (v2) at (-1,  1); 
        \coordinate (v3) at ( 1,  1); 
        \coordinate (v4) at ( 1, -1); 

        \coordinate (ext_v2) at (-2,  1);
        \coordinate (ext_v1) at (-2, -1);
        \coordinate (ext_v3) at ( 2,  1);
        \coordinate (ext_v4) at ( 2, -1);

        \draw[thick, dashed] (v1) -- (v2) node[midway, left] {$S^-$};

        \draw (v2) -- (v3) node[midway, below] {$E_L'$};
        \draw (v3) -- (v4) node[pos=0.2, right] {$L_L'$};
         \draw (v3) -- (v4) node[pos=0.8, right] {$L_R'$};
         \draw (v3) -- (v4) node[pos=0.5] {$\boldsymbol{\times}$};
        \draw (v4) -- (v1) node[midway, above] {$N_R$};

        \draw (ext_v2) -- (v2) node[pos=0, left] {$L_L(=\nu_L)$} node[pos=0.65, above right] {$y_1 P_L$};
        \draw (ext_v1) -- (v1) node[pos=0, left] {$l_R^-$} node[pos=0.65, below right] {$y_2 P_R$};
        \draw[thick, decorate, decoration={snake, amplitude=2pt, segment length=6pt}] (v3) -- (ext_v3) node[pos=0.3, above left] {$g_2 P_L$} node[pos=1, right] {$W_{\mu}^-$};
        \draw[thick, dashed] (v4) -- (ext_v4) node[pos=0.3, below left] {$y_3'^{ \dagger} P_L$} node[pos=1, right] {$H$};

        \path[postaction={decorate, decoration={markings, mark=at position 0.7 with {\arrow[scale=1.5]{latex}}}}] (ext_v2) -- (v2);

 \path[postaction={decorate, decoration={markings, mark=at position 0.6 with {\arrow[scale=1.5]{latex}}}}] (ext_v1) -- (v1);
 \path[postaction={decorate, decoration={markings, mark=at position 0.7 with {\arrow[scale=1.5]{latex}}}}] (v3) -- (v2);
 
   \path[postaction={decorate, decoration={markings, mark=at position 0.6 with {\arrow[scale=1.5]{latex}}}}] (v4) -- (v1);
   \path[postaction={decorate, decoration={markings, mark=at position 0.88 with {\arrow[scale=1.5]{latex}}}}] (v4) -- (v3);
   \path[postaction={decorate, decoration={markings, mark=at position 0.28 with {\arrow[scale=1.5]{latex}}}}] (v4) -- (v3);
   \path[postaction={decorate, decoration={markings, mark=at position 0.55 with {\arrow[scale=1.5]{latex}}}}]  (v1) -- (v2);
\path[postaction={decorate, decoration={markings, mark=at position 0.7 with {\arrow[scale=1.5]{latex}}}}] (v4) -- (ext_v4);

    \end{tikzpicture}
    \caption{Feynman diagram generating the operator $\mathcal{O}_{fW} = (\overline{L_L} \sigma^{\mu\nu} \tau^a l_R) H W_{\mu\nu}^a$.}
    \label{fig:square_loop_diagram4}
\end{figure}
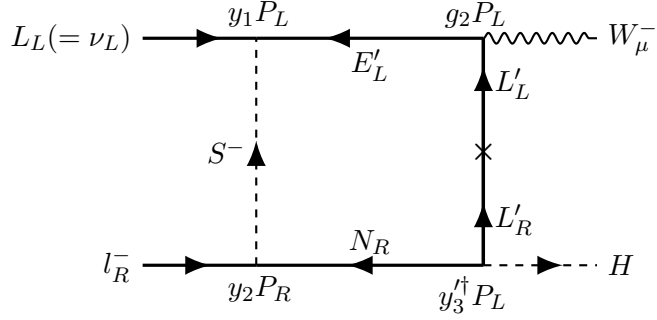

Using this interaction, the Feynman diagrams for the operator $\mathcal{O}_{fW} = (\overline{L_L} \sigma^{\mu\nu} \tau^a l_R) H W_{\mu\nu}^a$ are given in Figure~\ref{fig:square_loop_diagram3} and \ref{fig:square_loop_diagram4}.

Similar to the quark interactions, their coefficients are found as 
\begin{align}
\frac{g_2 y_1y_2 y_3}{(4\pi)^2 M_{L'}^2} F^L_1\left( \frac{m_S^2}{M_{L'}^2},  \frac{M_R^2}{M_{L'}^2}\right),\quad
\frac{g_2 y_1y_2 y'^\dag_3}{(4\pi)^2 M_{L'}^2} F^L_2\left( \frac{m_S^2}{M_{L'}^2},  \frac{M_R^2}{M_{L'}^2}\right),
\end{align}
where $m_S$ is the mass eigenvalue of $S^-$, and the loop functions $F^L_{1,2}={\cal O}(1)$ arise from  Figure~\ref{fig:square_loop_diagram3}, respectively.
%
While we have presented a concrete UV completion based on a scalar boson and fermion extension to illustrate the origin of the dimension-6 operators, a detailed phenomenological analysis of this specific model goes beyond the primary scope of the present work. In particular, the rigorous one-loop matching calculation and the quantitative evaluation of the resulting entanglement signatures are deferred to our forthcoming study.

We build a binned $\chi^2$ test statistic that incorporates systematic experimental uncertainties $\sigma_{\text{syst}}$ \cite{ATLAS:2018jlh} and statistical uncertainties (assuming an integrated luminosity corresponding to the High-Luminosity LHC, $\mathcal{L} = 3000\text{ fb}^{-1}$) in order to obtain projected $95\%$ CL exclusion bounds:
\bea
\chi^2(c_i) = \sum_{j} \frac{\left( \mathcal{O}_j^{\text{SM}+\text{EFT}}(c_i) - \mathcal{O}_j^{\text{SM}} \right)^2}{\sigma_{\text{stat},j}^2 + \sigma_{\text{syst},j}^2},
\eea
where the expected observable in the $j$-th kinematic and angular bin is indicated by $\mathcal{O}_j$. In our entanglement-based study, $\mathcal{O}_j$ denotes either the Von Neumann entropy perturbation $\delta S_{\text{VN}}$ or the binned linear entropy variation $\Delta S_L^{(i)}(s, \theta)$. The unpolarized differential cross-section $d\sigma/dt$ \cite{Aguilar-Saavedra:2022uye} is represented by $\mathcal{O}_j$ for conventional analyses.

The asymptotic sensitivity to effective Wilson coefficients is determined by the choice of observable, as demonstrated by evaluating the $\chi^2$ distribution for a single-parameter fit. Standard unpolarized cross-sections depend quadratically on the EFT contribution ($\propto 1/\Lambda^4$), which limits their exclusion reach in kinematic areas where chiral or angular symmetries cause the linear interference term to vanish \cite{Pomarol:2013zra,Peskin:1995ev}. On the other hand, entanglement measurements like $S_L$ and $S_{\text{VN}}$ maintain the leading linear $1/\Lambda^2$ scaling \cite{NielsenChuang2010} because they directly probe the phase coherence of the unsummed helicity density matrix. Analytically, this causes the parametric sensitivity to be driven by the first-order interference term at $\mathcal{O}(\Lambda^{-2})$ instead of being suppressed at order $\mathcal{O}(\Lambda^{-4})$. In high-energy invariant mass tails ($\sqrt{s} > 1\text{ TeV}$), where standard unpolarized rate measurements are otherwise limited by quadratic background dominance and symmetry cancellations, this structural recovery guaranties that entanglement-based observables retain strong sensitivity. 

\section{Experimental Realizability and Measurement Protocols}
Strong experimental procedures for reconstructing final-state density matrices are necessary to convert quantum entanglement metrics from a theoretical framework into a feasible collider observable \cite{Afik:2020onf,AfriatSelleri1999}. High-energy colliders (like the High-Luminosity LHC or future lepton colliders) rely on angular decay distributions, polarization asymmetries, and spin analyzers to infer density matrix elements because direct measurement of individual helicity amplitudes is not experimentally accessible \cite{Aguilar-Saavedra:2022uye}.

\subsection{Density Matrix Reconstruction via Spin Analyzers}
The decay products of unstable particles created in high-energy collisions, including weak gauge bosons, tau leptons, or top quarks, serve as ideal, integrated spin analyzers. The polarization vector and spin correlation matrix of the parent particles have a linear correlation with the angular distribution of these decay products \cite{Peskin:1995ev}:
\bea
\frac{1}{\sigma} \frac{d\sigma}{d\Omega_a d\Omega_b} = \frac{1}{4\pi^2} \sum_{i,j=1}^3 B_i B_j \omega_i^a \omega_j^b,
\eea
where $B_i$ stands for the elements of the spin polarization vector and $\omega_i^{a,b}$ for the angular distributions of the decay products in their corresponding rest frames. Experimental collaborators can reconstruct the entire two-particle spin density matrix $\rho_{\text{final}}$ by measuring the joint angular distributions of decay products across several orthogonal axis without needing to fully reconstruct each microscopic vertex \cite{NielsenChuang2010}.

\subsection{Statistical Uncertainties and High-Luminosity Projections}
There are particular statistical difficulties when extracting quantum entanglement measures from binned density matrix elements. The number of recorded events $N_{\text{ev}}$ within each kinematic and angular bin $(\Delta s, \Delta\cos\theta)$ \cite{ATLAS:2018jlh} directly affects the precision of linear entropy variations $\Delta S_L$ and Von Neumann entropy shifts $\delta S_{\text{vN}}$:
\bea
\sigma(\Delta S_L) \sim \frac{1}{\sqrt{N_{\text{ev}}}}.
\eea
With an integrated luminosity of $\mathcal{L} = 3000\text{ fb}^{-1}$ at the HL-LHC, high-statistics channels (like vector boson scattering or $t\bar{t}$ production) produce enough event counts in high-invariant-mass tails ($\sqrt{s} > 1\text{ TeV}$) to keep statistical fluctuations well below the estimated linear SMEFT signal size $v^2/\Lambda^2$ \cite{Pomarol:2013zra}.

\subsection{Systematic Error Management and Detector Effects}
Rigorous control over experimental systematics, such as jet energy scale errors, lepton reconstruction efficiency, and background contamination from misidentified objects, is necessary for the precise extraction of entanglement observables \cite{Campbell:2006wx}. Detector acceptance and finite resolution can dilute purity metrics because entanglement measures depend on off-diagonal coherence elements, persistently biasing entropy estimates toward separable states. Before calculating $S_L$ and $S_{\text{VN}}$ \cite{AfriatSelleri1999, ATLAS:2018jlh}, contemporary analyses use unfolding techniques (such as iterative Bayesian unfolding or machine learning-based matrix inversion) to map detector-level distributions back to true particle-level density matrices.

\section{Conclusion and results}
By establishing a rigorous theoretical framework that connects effective field theory and quantum information science, this work shows that quantum entanglement metrics, namely linear and Von Neumann entropies, are high-sensitivity probes for physics beyond the Standard Model at high-energy colliders. Our method maintains the phase, sign, and angular interference structure between dimension-six effective operators and Standard Model baselines by building reduced density matrices $\rho_A$ from unsummed helicity scattering amplitudes. Entanglement-based observables inherit a direct linear dependency on the Wilson coefficients, $c_i/\Lambda^2$, in contrast to conventional unpolarized total cross-sections that suffer from severe angular cancellations and quadratic $1/\Lambda^4$ suppression. Differential binning allows for the isolation of particular chiral interference patterns since this linear scaling is maintained throughout phase-space and angular distributions. Entanglement diagnostics greatly outperform traditional cross-section limitations, as demonstrated by phenomenological case studies of electroweak dipole and gauge-portal operators. According to projected sensitivity analyses for the High-Luminosity LHC ($\mathcal{L} = 3000\text{ fb}^{-1}$), exclusion reach on $c_i/\Lambda^2$ is improved by linear entropy variations $\Delta S_L$ and Von Neumann entropy shifts $\delta S_{\text{vN}}$, especially in high-invariant-mass tails ($\sqrt{s} > 1\text{ TeV}$). Moreover, our protocols show that robust density matrix reconstruction is possible without requiring full microscopic vertex monitoring by using conventional decay products, such as top quarks, tau leptons, and weak bosons, as built-in spin analysers. In the future, the techniques proposed here provide a number of intriguing directions for quantum information and collider phenomenology study. A logical next step would be to extend the density matrix formalism beyond bipartite systems to multipartite entanglement measurements, like tangle or residual entanglement in multi-boson production channels. Furthermore, by incorporating neural network-based unfolding approaches, systematic errors from acceptance cuts and finite resolution will be reduced by directly mapping detector-level kinematic distributions to genuine particle-level spin density matrices. Lastly, using these quantum diagnostics at future lepton colliders, such the FCC-ee or ILC, will take advantage of enhanced beam polarization and clean starting states to significantly reduce background contamination and push accuracy of SMEFT bounds. In the end, quantum information tools are transformed from conceptual underpinnings into potent discovery tools for fundamental particle physics by framing quantum entanglement as an experimental observable.

 \section{Acknowledgement}
 H.O. is supported by Zhongyuan Talent (Talent Recruitment Series) Foreign Experts Project.

\bibliography{References}

\providecommand{\href}[2]{#2}\begingroup\begin{thebibliography}{10}

\bibitem{Weinberg:1995mt}
S.~Weinberg, \textit{{The Quantum theory of fields. Vol. 1: Foundations}}.
\newblock Cambridge University Press, 6, 2005,
  \href{https://doi.org/10.1017/CBO9781139644167}{10.1017/CBO9781139644167}.

\bibitem{Jangid:2023hzi}
S.~Jangid, K.~I. Nagao and H.~Okada, \textit{{Neutrino mass model and dark
  matter with Y=0 inert triplet scalar}},
  \href{https://doi.org/10.1016/j.physletb.2024.139030}{\textit{Phys. Lett. B}
  {\bfseries 858} (2024) 139030},
  [\href{https://arxiv.org/abs/2311.08901}{{\ttfamily 2311.08901}}].

\bibitem{Biswas:2025woq}
A.~Biswas, S.~Jangid and S.~C. Park, \textit{{Investigating two-zero texture in
  the light of gauged Type-II seesaw}},
  \href{https://doi.org/10.1016/j.physletb.2025.139786}{\textit{Phys. Lett. B}
  {\bfseries 868} (2025) 139786},
  [\href{https://arxiv.org/abs/2504.10312}{{\ttfamily 2504.10312}}].

\bibitem{Jangid:2023jya}
S.~Jangid and H.~Okada, \textit{{Exploring CP-violation in Y=0 inert triplet
  with real singlet}},
  \href{https://doi.org/10.1103/PhysRevD.108.055025}{\textit{Phys. Rev. D}
  {\bfseries 108} (2023) 055025},
  [\href{https://arxiv.org/abs/2304.13325}{{\ttfamily 2304.13325}}].

\bibitem{Buchmuller:1985jz}
W.~Buchmuller and D.~Wyler, \textit{{Effective Lagrangian Analysis of New
  Interactions and Flavor Conservation}},
  \href{https://doi.org/10.1016/0550-3213(86)90262-2}{\textit{Nucl. Phys. B}
  {\bfseries 268} (1986) 621--653}.

\bibitem{Grzadkowski:2010es}
B.~Grzadkowski, M.~Iskrzynski, M.~Misiak and J.~Rosiek, \textit{{Dimension-Six
  Terms in the Standard Model Lagrangian}},
  \href{https://doi.org/10.1007/JHEP10(2010)085}{\textit{JHEP} {\bfseries 10}
  (2010) 085}, [\href{https://arxiv.org/abs/1008.4884}{{\ttfamily 1008.4884}}].

\bibitem{Afik:2020onf}
Y.~Afik and J.~R.~M. de~Nova, \textit{{Entanglement and quantum tomography with
  top quarks at the LHC}},
  \href{https://doi.org/10.1140/epjp/s13360-021-01902-1}{\textit{Eur. Phys. J.
  Plus} {\bfseries 136} (2021) 907},
  [\href{https://arxiv.org/abs/2003.02280}{{\ttfamily 2003.02280}}].

\bibitem{Aguilar-Saavedra:2022uye}
J.~A. Aguilar-Saavedra and J.~A. Casas, \textit{{Improved tests of entanglement
  and Bell inequalities with LHC tops}},
  \href{https://doi.org/10.1140/epjc/s10052-022-10630-4}{\textit{Eur. Phys. J.
  C} {\bfseries 82} (2022) 666},
  [\href{https://arxiv.org/abs/2205.00542}{{\ttfamily 2205.00542}}].

\bibitem{ATLAS:2023fsd}
{\scshape ATLAS} collaboration, G.~Aad et~al., \textit{{Observation of quantum
  entanglement with top quarks at the ATLAS detector}},
  \href{https://doi.org/10.1038/s41586-024-07824-z}{\textit{Nature} {\bfseries
  633} (2024) 542--547}, [\href{https://arxiv.org/abs/2311.07288}{{\ttfamily
  2311.07288}}].

\bibitem{CMS:2024pts}
{\scshape CMS} collaboration, A.~Hayrapetyan et~al., \textit{{Observation of
  quantum entanglement in top quark pair production in
  proton{\textendash}proton collisions at $\sqrt{s} = 13$ TeV}},
  \href{https://doi.org/10.1088/1361-6633/ad7e4d}{\textit{Rept. Prog. Phys.}
  {\bfseries 87} (2024) 117801},
  [\href{https://arxiv.org/abs/2406.03976}{{\ttfamily 2406.03976}}].

\bibitem{Weinberg:1967kj}
S.~Weinberg, \textit{{Precise relations between the spectra of vector and axial
  vector mesons}},
  \href{https://doi.org/10.1103/PhysRevLett.18.507}{\textit{Phys. Rev. Lett.}
  {\bfseries 18} (1967) 507--509}.

\bibitem{Georgi:1993mps}
H.~Georgi, \textit{{Effective field theory}},
  \href{https://doi.org/10.1146/annurev.ns.43.120193.001233}{\textit{Ann. Rev.
  Nucl. Part. Sci.} {\bfseries 43} (1993) 209--252}.

\bibitem{Mangano:1990by}
M.~L. Mangano and S.~J. Parke, \textit{{Multiparton Amplitudes in Gauge
  Theories}}, \href{https://doi.org/10.1016/0370-1573(91)90091-Y}{\textit{Phys.
  Rept.} {\bfseries 200} (1991) 301--367},
  [\href{https://arxiv.org/abs/hep-th/0509223}{{\ttfamily hep-th/0509223}}].

\bibitem{Dixon:1996wi}
L.~J. Dixon, \textit{{Calculating scattering amplitudes efficiently}},  in
  \textit{{Theoretical Advanced Study Institute in Elementary Particle Physics
  (TASI 95): QCD and Beyond}}, pp.~539--584, 1, 1996,
  \href{https://arxiv.org/abs/hep-ph/9601359}{{\ttfamily hep-ph/9601359}}.

\bibitem{LHCHiggsCrossSectionWorkingGroup:2011wcg}
{\scshape LHC Higgs Cross Section Working Group} collaboration, S.~Dittmaier
  et~al., \textit{{Handbook of LHC Higgs Cross Sections: 1. Inclusive
  Observables}},  \href{https://arxiv.org/abs/1101.0593}{{\ttfamily
  1101.0593}}.

\bibitem{Pomarol:2013zra}
A.~Pomarol and F.~Riva, \textit{{Towards the Ultimate SM Fit to Close in on
  Higgs Physics}}, \href{https://doi.org/10.1007/JHEP01(2014)151}{\textit{JHEP}
  {\bfseries 01} (2014) 151},
  [\href{https://arxiv.org/abs/1308.2803}{{\ttfamily 1308.2803}}].

\bibitem{Severi:2021cnj}
C.~Severi, C.~D.~E. Boschi, F.~Maltoni and M.~Sioli, \textit{{Quantum tops at
  the LHC: from entanglement to Bell inequalities}},
  \href{https://doi.org/10.1140/epjc/s10052-022-10245-9}{\textit{Eur. Phys. J.
  C} {\bfseries 82} (2022) 285},
  [\href{https://arxiv.org/abs/2110.10112}{{\ttfamily 2110.10112}}].

\bibitem{vonNeumann1955}
J.~von Neumann, \textit{Mathematical Foundations of Quantum Mechanics}.
\newblock Princeton University Press, Princeton, NJ, 1955,
  \href{https://doi.org/10.2307/j.ctt1wq8zhp}{10.2307/j.ctt1wq8zhp}.

\bibitem{NielsenChuang2010}
M.~A. Nielsen and I.~L. Chuang, \textit{Quantum Computation and Quantum
  Information}.
\newblock Cambridge University Press, Cambridge, 10th anniversary~ed., 2010.

\bibitem{Bernreuther2008}
W.~Bernreuther, \textit{Perturbative {QCD} and {CP} violation in heavy quark
  decays and colliders},  in \textit{Top Quark Physics} (W.~Hollik, ed.),
  vol.~742 of \textit{Lecture Notes in Physics}, pp.~39--82.
\newblock Springer, Berlin, Heidelberg, 2008.
\newblock \href{https://doi.org/10.1007/978-3-540-74783-5_2}{DOI}.

\bibitem{AfriatSelleri1999}
A.~Afriat and F.~Selleri, \textit{The Einstein, Podolsky, and Rosen Paradox in
  Atomic, Nuclear, and Particle Physics}.
\newblock Springer, New York, NY, 1999,
  \href{https://doi.org/10.1007/978-1-4757-3081-3}{10.1007/978-1-4757-3081-3}.

\bibitem{HiggsWorkingGroup:2004ndr}
{\scshape Higgs Working Group} collaboration, K.~A. Assamagan et~al.,
  \textit{{The Higgs working group: Summary report 2003}},  in \textit{{3rd Les
  Houches Workshop on Physics at TeV Colliders}}, pp.~1--169, 6, 2004,
  \href{https://arxiv.org/abs/hep-ph/0406152}{{\ttfamily hep-ph/0406152}}.

\bibitem{Campbell:2006wx}
J.~M. Campbell, J.~W. Huston and W.~J. Stirling, \textit{{Hard Interactions of
  Quarks and Gluons: A Primer for LHC Physics}},
  \href{https://doi.org/10.1088/0034-4885/70/1/R02}{\textit{Rept. Prog. Phys.}
  {\bfseries 70} (2007) 89},
  [\href{https://arxiv.org/abs/hep-ph/0611148}{{\ttfamily hep-ph/0611148}}].

\bibitem{Peskin:1995ev}
M.~E. Peskin and D.~V. Schroeder, \textit{{An Introduction to quantum field
  theory}}.
\newblock Addison-Wesley, Reading, USA, 1995,
  \href{https://doi.org/10.1201/9780429503559}{10.1201/9780429503559}.

\bibitem{Kleiss:1985yh}
R.~Kleiss and W.~J. Stirling, \textit{{Spinor Techniques for Calculating p
  anti-p ---{\ensuremath{>}} W+- / Z0 + Jets}},
  \href{https://doi.org/10.1016/0550-3213(85)90285-8}{\textit{Nucl. Phys. B}
  {\bfseries 262} (1985) 235--262}.

\bibitem{Dedes:2021abc}
A.~Dedes and K.~Mantzaropoulos, \textit{{Universal scalar leptoquark action for
  matching}}, \href{https://doi.org/10.1007/JHEP11(2021)166}{\textit{JHEP}
  {\bfseries 11} (2021) 166},
  [\href{https://arxiv.org/abs/2108.10055}{{\ttfamily 2108.10055}}].

\bibitem{Chala:2020vqp}
M.~Chala and A.~Titov, \textit{{One-loop matching in the SMEFT extended with a
  sterile neutrino}},
  \href{https://doi.org/10.1007/JHEP05(2020)139}{\textit{JHEP} {\bfseries 05}
  (2020) 139}, [\href{https://arxiv.org/abs/2001.07732}{{\ttfamily
  2001.07732}}].

\bibitem{Gu:2024wrk}
J.~Gu, J.~Guo and X.-Z. Tan, \textit{{Tracing the bottom electroweak dipole
  operators at future lepton colliders}},
  \href{https://doi.org/10.1007/JHEP06(2025)035}{\textit{JHEP} {\bfseries 06}
  (2025) 035}, [\href{https://arxiv.org/abs/2410.05398}{{\ttfamily
  2410.05398}}].

\bibitem{deBlas:2017xtg}
J.~de~Blas, J.~C. Criado, M.~Perez-Victoria and J.~Santiago, \textit{{Effective
  description of general extensions of the Standard Model: the complete
  tree-level dictionary}},
  \href{https://doi.org/10.1007/JHEP03(2018)109}{\textit{JHEP} {\bfseries 03}
  (2018) 109}, [\href{https://arxiv.org/abs/1711.10391}{{\ttfamily
  1711.10391}}].

\bibitem{Henning:2014wua}
B.~Henning, X.~Lu and H.~Murayama, \textit{{How to use the Standard Model
  effective field theory}},
  \href{https://doi.org/10.1007/JHEP01(2016)023}{\textit{JHEP} {\bfseries 01}
  (2016) 023}, [\href{https://arxiv.org/abs/1412.1837}{{\ttfamily 1412.1837}}].

\bibitem{ATLAS:2018jlh}
{\scshape ATLAS} collaboration, \textit{{Projections for measurements of Higgs
  boson cross sections, branching ratios, coupling parameters and mass with the
  ATLAS detector at the HL-LHC}}, .

\end{thebibliography}\endgroup
\bibliographystyle{Ref}
\end{document}